# Is the Goldman-Hodgkin-Katz equation universally true?

Yoav Green[*]

Department of Mechanical Engineering, Ben-Gurion University of the Negev, Beer-Sheva 8410501, Israel

[*] Email: yoavgreen@bgu.ac.il
ORCID number: 0000-0002-0809-6575

**Abstract.**

**No.** Eighty years ago, the two seminal works by Goldman [Journal of General Physiology 27, 37 (1943)] and by Hodgkin-Katz [The Journal of Physiology 108, 37 (1949)] derived the foundational framework for interpreting electro-physiological measurements in what is commonly termed the Goldman-Hodgkin-Katz (GHK) theory for the membrane potential. Both seminal papers postulate a constant/uniform electric field within the ion channel. Using a uniform electric field allows for a simple, straightforward calculation of the ionic fluxes and the transmembrane potential, which yields the famous GHK potential. The use of this framework is so widely accepted that one can find a plethora of works that no longer cite the original works, or many books and reviews that discuss GHK. Thus, GHK has perhaps become the "universal" and indisputable descriptor of the underlying physics and biology. In a recent set of works [Phys. Rev. Lett. 134, 228401 (2025) and Phys. Rev. E 111, 064408 (2025)], we revisited GHK and its assumption of a uniform field. Non-approximated numerical simulations showed that the electric field is not always uniform. To understand this discrepancy, it is important to understand that the governing equations can be solved using two different approaches: the GHK approach of assuming a constant electric field or postulating that the system is electroneutral (an important physical trait that GHK does/can not satisfy). Each approach yields drastically different non-commutative results. The purpose of this report is to provide a non-mathematical summary of the results and to inform the broader community that GHK is not as universal as previously thought. We will discuss these two newer works, with some emphasis on how they can potentially revolutionize the interpretation of electro-physiological measurements. Importantly, we show that this new framework, which utilizes the mathematical tools developed by the electrodialysis community, also serves as a bridge linking the electrophysiology community and the electrodialysis community.

**Key points.**

The Goldman-Hodgkin-Katz equation has been the paradigm theory for ion transport in physiological systems for the past century. Here, we show that this theory is internally inconsistent. In its place, we provide a new, internally consistent theory that includes expressions for all the main characteristics of ion transport in ion channels. Numerical simulations substantiate our theory. This new theory relates/connects GHK to the field of reverse-electrodialysis (RED), and provides a remarkably robust framework for interpreting (and reinterpreting) ion transport experiments in any charge-selective system.

# 1. Introduction

Approximately eighty years ago, the Goldman-Hodgkin-Katz (GHK) equation for the membrane potential was derived by Goldman (Goldman, 1943) and by Hodgkin-Katz (Hodgkin & Katz, 1949) in their two seminal works that have since then become the cornerstone of the electrophysiology and ion channel communities, but can even be found in the desalination community (Zhang et al., 2025). At the time of the writing of this report, based on Web of Science[1], these works of Goldman (Goldman, 1943) and by Hodgkin-Katz (Hodgkin & Katz, 1949) have been cited approximately 2,600 and 3200 times, respectively. GHK is so widespread that one does not even need to reference a textbook (Hille, 2001; Koch, 2004; Kandel et al., 2012) and reviews (Alvarez & Latorre, 2017), or even the original works, to be understood without loss of understanding for the reader.

Given the vast number of experimental works that have utilized GHK, it is impossible to address a specific work or even a large body of them. This is not the goal of this work (although implications for experiments will be discussed later). Rather, our goal has more of a "mathematical" nature. Our goal is to ascertain if the GHK model is a distinct and unique solution to the governing equations. We will show that it is not. However, we assure the reader that we will not invoke much mathematics in this work. Rather, we will focus here on the outcomes that were recently reported in (Green, 2025a, 2025b). The interested reader will be able to find a detailed (mathematical) discussion in these works.

Like with any theoretical mathematical model, one needs to appreciate that during the derivation process, certain assumptions are inserted into the model. Some of these assumptions are justifiable on physical grounds, while others are placed to simplify the derivation process further. Given a set of equations, assumption $X_1$ can lead to result $Y_1$, while a different assumption, $X_2$, can lead to a result $Y_2$ such that $Y_2 \neq Y_1$. GHK is no different. GHK theory is derived from a set of equations, wherein it is assumed that the electric field is uniform throughout the channel. However, what happens if the exact set of equations is solved while the assumption of the uniform electric field is replaced with a different assumption? Will the response change in a non-trivial manner? The answer to these questions is "Yes! The response changes drastically!"

For the sake of completion, in the main text, we provide a non-mathematical overview of the underlying physical model used to derive GHK. In Appendix A, we provide the governing equations and boundary conditions. These are the same equations used by Goldman (Goldman, 1943) and Hodgkin-Katz (Hodgkin & Katz, 1949) in their seminal works, with the exception that we focus on two species. The Supplemental Information of (Green, 2025b) provides a table detailing the differences in notations and terminology between our work and theirs. The remaining information regarding the derivation and numerical simulations can be found in our recent work (Green, 2025b).

# 2. Model overview

Electrophysiologists consider transport of ions through an ion channel embedded in an insulating biological membrane, such as the bottom channel shown in Figure 1. However, this is not that different from the classical problem of transport across a simple nanochannel embedded into an insulating dielectric material (top channel shown in Figure 1). In fact, in GHK, it is assumed that transport is 1D (such that the geometry is uniform). As a result, transport through an ion channel and a nanochannel is essentially the same and should have identical responses. This statement is particularly true since steady-state and convectionless ion transport through any ion-selective system is governed by the same non-linear coupled Poisson-Nernst-Planck (PNP) equations (Goldman, 1943).

---

[1] Certain data included herein are derived from Clarivate™ (Web of Science™). © Clarivate 2025. All rights reserved.

At the two ends of the systems, one has asymmetric bulk concentrations, denoted by $c_{\text{left}}$ and $c_{\text{right}}$, and a potential drop $V$. Both the concentration gradient and the electric potential gradient will induce an electrical current density $i$. It is here that the interests of the electro-physiological community, and water desalination and energy harvesting communities differ: GHK is particularly interested in the voltage when $i = 0$ (i.e., $V_{i=0}$), while the desalination and energy harvesting communities, to which the author belongs, are more interested in the "harvestable" current when $V = 0$ (i.e., $i_{V=0}$). However, since $V$ and $i$ are interrelated, knowledge of one provides knowledge of the other.

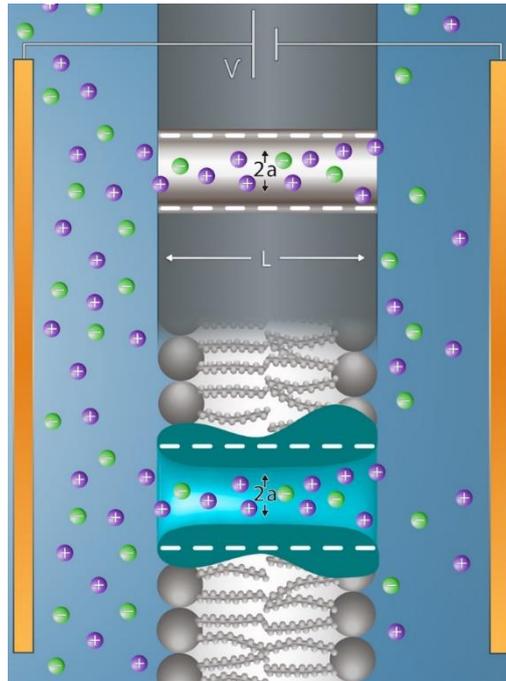

**Figure 1**. Schematic of two different nanofluidic systems that have been merged into one. On the bottom, we have an ion channel embedded into a biological membrane, which can be represented by the top nanopore that is embedded in a simple dielectric material. The biological membrane or dielectric material are completely isolating such that fluxes are allowed only through the channel. Both channels are charged with an embedded surface charge density (here, negative and denoted by white minus signs). The surface charge leads to a surplus of purple counterions (here, positive) over the green coions (here, negative). The channels are subjected to a combined voltage drop $V$ (defined positive from left to right) and asymmetric concentrations denoted by $c_{\text{left}}$ and $c_{\text{right}}$. In general, $c_{\text{left}}$ and $c_{\text{right}}$ are not equal. Here, we have depicted that $c_{\text{left}} > c_{\text{right}}$. However, the reverse situation is also allowable. Reprinted with permission from (Green, 2025a, 2025b). Copyright (2025) (American Physical Society).

For two species (one positive and one negative), the mathematical formulation of the PNP equations, subject to asymmetric concentrations and potential drop, is unambiguous (see Appendix A for the mathematical formulation and for a discussion on extending beyond two species). The solution, however, is not. As we have shown (Green, 2025a, 2025b) and will discuss here, the underlying assumptions and the inserted parameters can yield two non-commutative and distinctly different solutions. For example, the derived expressions for $V_{i=0}$ (or any transport characteristic) depends on the embedded assumptions used in the derivation.

The differences in the parameters and assumptions can be attributed to the field of inquiry. For example, the GHK community and the water desalination and energy-harvesting communities that utilize electrodialysis (ED) and reverse-electrodialysis (RED), respectively, hold different views on the importance of the surface charge density (denoted by the white minus signs embedded into the channels in Figure 1). Often, the effects of the surface charge density are transformed to a parameter that has units of concentration and is termed the fixed charges. We will denote the fixed charges by $\Sigma_s$. In GHK, it is explicitly assumed that the fixed charges are identically zero ($\Sigma_s = 0$). In contrast, in the realm of water desalination through electrodialysis (ED) and energy harvesting through reverse-electrodialysis (RED), the non-zero surface charge density is what breaks the symmetry of the transport of positive and negative ions, such that the limit of $\Sigma_s = 0$ is inconceivable. Nonetheless, the mathematical limit of $\Sigma_s = 0$ is not singular in ED/RED, and the predictions derived for ED/RED can be (and will be) tested at this limit.

One must ask the immediate question: if one takes $\Sigma_s = 0$ in ED/RED, do the results recapitulate the results of GHK? We will shortly show that the answer to this is "No". This "No" is also why the GHK equation is not as universal as previously thought. The reason that GHK and ED/RED have different results is not just whether or not they account for $\Sigma_s$, but more importantly, it is due to the different embedded assumptions on how to simplify and solve the governing equations (in particular, the Poisson equation for the electric potential).

GHK assumes that the electric potential has a constant electric field (or linear electric potential distribution). Once this assumption is asserted, electroneutrality of the space charge density is not satisfied (locally or even globally). ED/RED utilizes the exact opposite approach, wherein electroneutrality is locally (and, thus, globally) enforced, but the electric field remains to be determined. It is this methodology difference alone that is responsible for the different predictions and interpretations of these two communities. Appendix B provides a slightly more detailed mathematical discussion regarding this issue, and more details can be found in (Green, 2025a, 2025b).

### 3. GHK theory

For two species, one positive and one negative, with the respective diffusion coefficients $D_+$ and $D_-$, subjected to asymmetric bulk concentrations denoted by $c_{\text{left}}$ and $c_{\text{right}}$, the GHK membrane potential for an uncharged channel is given by (Goldman, 1943; Hodgkin & Katz, 1949)

$$V_{\text{GHK}} = V_{th} \ln\left(\frac{D_+ c_{\text{right}} + D_- c_{\text{left}}}{D_+ c_{\text{left}} + D_- c_{\text{right}}}\right). \tag{1}$$

where $V_{th} = RT/Fz$ is the thermal potential that depends on the universal gas constant, $R$, the absolute temperature, $T$, assumed to be uniform across the system, $F$ the Faraday constant, and the valency $z$ (which is assumed to be equal but of opposite signs for the positive and negative ions $z_\pm = \pm z$).

The derivation of Eq. (1) includes several embedded assumptions. In their Appendix, Hodgkin-Katz (Hodgkin & Katz, 1949) succinctly write[2]

*"The basic assumptions are (1) that ions in the membrane move under the influence of diffusion and the electric field in a manner which is essentially similar to that in free solution; (2) that the electric field may be regarded as constant throughout the membrane; (3) that the concentrations of ions at the edges of the membrane are directly*

---

[2] Hodgkin-Katz assumptions are equivalent with Goldman's. However Goldman's approach and line of reasoning are slightly more circular and complicated.

*proportional to those in the aqueous solutions bounding the membrane; and (4) that the membrane is homogeneous."*

Assumption (1) implies several outcomes. That the behavior in the channel itself is equal to that in "free solution" or in bulk implies that the channel itself is uncharged, leading to the fixed charges being zero ($\Sigma_s = 0$), such that it is tantamount to the removal of the surface charge density. Furthermore, based on their statement, it can be assumed that the diffusion coefficients inside and outside the membrane are the same – bulk diffusion coefficients. Assumption (2) repeats Goldman's assumption of a uniform electric field. Assumption (3) is an assumption that at the two edges of the system, the concentrations are $c_{\text{left}}$ and $c_{\text{right}}$. Finally, assumption (4) is that fluxes are uniform over the cross-section (equivalent to assuming 1D behavior). The assumption of 1D is also what leads to the equivalency of the ion-channel system to the simple nanochannel system.

One can often find Eq. (1) where ionic permeabilities or replace the diffusion coefficients. It is essential to note that when the permeabilities are used, they are considered to be fitting parameters. However, one needs to consider two caveats associated with using them as fitting parameters. First, it is not consistent with assumption (1), which assumes that properties are bulk properties. If the diffusion coefficients outside the channels are known, how are they suddenly different within the channel? Second, it should be remembered that with enough fitting parameters, one can always fit the data.

## 4. Why does GHK fail?

Before we go into a detailed discussion of the uniform electric field assumption, it is first worthwhile to demonstrate its failure. Figure 2 presents Comsol finite element simulations of Poisson-Nernst-Planck equations subject to asymmetric bulk concentration ($c_{\text{left}} \neq c_{\text{right}}$) and subject to a zero-current [the simulation details can be found in (Green, 2025b)]. It can be observed that the electric potential is not a linear profile (with a constant electric field), but rather the potential has a logarithmic profile (Green, 2025b)! If the electric field is not constant, then all outcomes assuming it are no longer correct. In particular, the famous GHK membrane potential [Eq. (1)] is not as universal as previously assumed, a major point of this report.

The reason that GHK deviates from numerical simulations can be attributed to the approach in solving for the governing equation (the Poisson equation) for the electric potential. The Poisson equation is the differential form of Gauss's law, which states that electric charges create electric fields wherein the electric field moves the charges, which influence how the electric field redistributes itself, etc. There is a two-way coupling between the electric field and the electric charges. In both Goldman's approach and our approach, we remove this two-way coupling, but we do so differently.

In the following, we shall discuss these differences briefly and without an in-depth [see Appendix B and/or (Green, 2025b) for a detailed mathematical discussion]. Goldman (Goldman, 1943) assumes that the effects of the charges on the electric field are negligible, such that the electric field in the channel is equal to that of a region without charge – this is the uniform electric field assumption [equivalent to the linear potential drop]. This uniform electric field is then inserted into the Nernst-Planck equations, yielding the concentrations and fluxes. However, a self-consistency check shows the difference between the positive and negative ions (which yields the space charge density) is never zero locally or even globally [this result is shown in Fig. 5(a) of (Green, 2025b)]. If there is a space-charge density, it would create an internal field, which would result in the deviation from the uniform field assumption.

In contrast, the second approach used by us is to assume that the local change in the electric field does not strongly influence the redistribution of the charges. This leads to the assumption that the electrolyte is electroneutral, which in turn leads to a relation between the positive and negative ionic concentration (and the fixed charges). Thereafter, the Nernst-Planck equations are solved for the remaining concentration and the electric potential. The resultant expression for the potential with zero surface charge ($\Sigma_s = 0$) is logarithmic – this is the result shown in Figure 2. The advantage of this electroneutral approach is that the system is locally and globally electroneutral, and one does not need to worry about how local charges might attract charges from outside of the system. Importantly, this approach allows us to remove the overly restrictive assumption of $\Sigma_s = 0$. This new approach can be extended to hold for arbitrary values of $\Sigma_s$, which cannot be calculated within GHK, and yet is an essential component of RED systems.

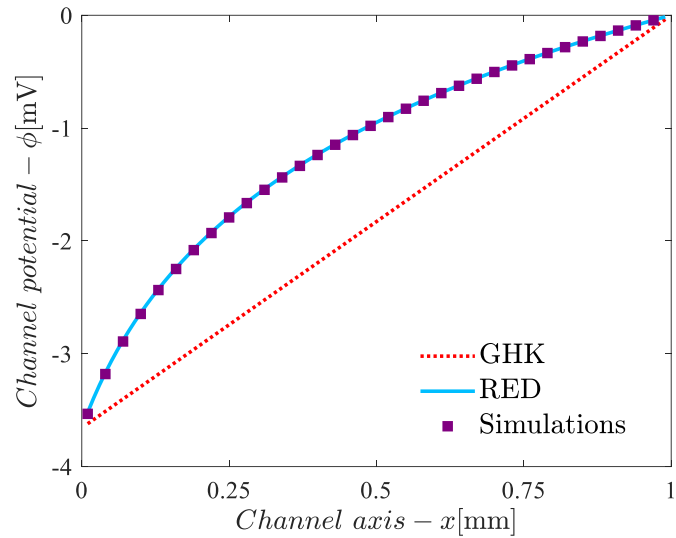

**Figure 2**. Comparison of the potential distribution, $\phi(x)$, showing that numerical simulations fall on the logarithmic solution predicted by the electroneutral/RED model, and not the linear potential (constant field) predicted by GHK. Adapted/reprinted with permission from (Green, 2025a, 2025b). Copyright (2025) (American Physical Society).

### 5. Electroneutral model

Having established that GHK is not a unique and universal solution, an alternative theory is needed. In this alternative theory, we retain Hodgkin-Katz (Hodgkin & Katz, 1949) assumptions (3) and (4). We can now also consider a less restrictive assumption (1): we will no longer require that the fixed charges are identically zero, rather $\Sigma_s$ is arbitrary. However, for the sake of comparison, the limit $\Sigma_s = 0$ will also be considered, and it will be shown to differ from GHK. We will also not treat the diffusion coefficients as a fitting parameter (they are given by bulk properties). The sole difference between the two theories will be the replacing assumption (2), of a uniform electric field, with the assumption of local electroneutrality.

The resultant expression for the membrane potential for arbitrary values of $D_+$, $D_-$, $\Sigma_s$, $c_{\text{left}}$, and $c_{\text{right}}$ is given by (Galama et al., 2016; Lavi & Green, 2024; Green, 2025a, 2025b)

$$V_{i=0} = V_D + V_{\text{no-D}}, \tag{2}$$

wherein

$$V_{\text{no-D}} = V_{th}\left[\ln\left(\frac{c_{\text{right}}}{c_{\text{left}}}\right) - \ln\left(\frac{\sqrt{\Sigma_s^2 + 4c_{\text{right}}^2} + \Sigma_s}{\sqrt{\Sigma_s^2 + 4c_{\text{left}}^2} + \Sigma_s}\right)\right], \tag{3}$$

$$V_D = V_{th}\frac{D_+ - D_-}{D_+ + D_-}\ln\frac{Q_{\text{right}}}{Q_{\text{left}}}, \tag{4}$$

$$Q_{k=\text{left,right}} = (D_+ - D_-)\Sigma_s + (D_+ + D_-)\sqrt{\Sigma_s^2 + 4c_k^2}. \tag{5}$$

Here, $V_{\text{no-D}}$ is the potential drop that is independent of the diffusion coefficient, while $V_D$ is the additional contribution due to the asymmetry of the diffusion coefficients ($D_+ \neq D_-$).

For the case of zero fixed charges ($\Sigma_s = 0$), corresponding to the GHK model [Eq. (1)], Eq. (3) simplifies to $V_{\text{no-D}} = 0$, while Eqs. (4)-(5) are also further simplified such that the membrane potential given by Eq. (2) simplifies to what is termed the Henderson equation (Galama et al., 2016)

$$V_{\text{Henderson}} = V_{th} \frac{D_+ - D_-}{D_+ + D_-} \ln\left(\frac{c_{\text{right}}}{c_{\text{left}}}\right). \tag{6}$$

This equation can also be found in Goldman's original work (Goldman, 1943), where it is all but ignored [this issue, too, is discussed in (Green, 2025b)]. One can already note that Eq. (1) and Eq. (6) have different forms and provide substantially different predictions (Green, 2025b) for $V_{i=0}$ (except for KCl, where $D_+ = D_-$).

It can also be shown that in the limit of low concentrations [or when $\Sigma_s \gg \max(c_{\text{left}}, c_{\text{right}})$], which is the exact opposite limit of vanishing selectivity when $\Sigma_s = 0$ or $\Sigma_s \ll \min(c_{\text{left}}, c_{\text{right}})$, Eq. (2) reduces to the Nernst potential

$$V_{\text{Nernst}} = V_{th} \ln\left(\frac{c_{\text{right}}}{c_{\text{left}}}\right). \tag{7}$$

It should also be noted that within GHK theory, one does not have a self-consistent expression that relates Eqs. (1) and (7) for all concentrations (and/or all $\Sigma_s$).

Figure 3 summarizes these results for three salts: HCl, KCl, and NaCl, corresponding to the three respective cases of $D_+ > D_-$, $D_+ = D_-$, and $D_+ < D_-$. We observe the remarkable correspondence between the new RED model [Eq. (2)] and numerical simulation for all concentrations, including very high [Eq. (6)] and very low [Eq. (7)] concentrations denoted by the dotted orange lines and dashed black, respectively, in Figure 3. One can immediately note that Eq. (1) [denoted by the dashed-dotted blue lines in Figure 3] does not have good correspondence. Depending on the parameters, this difference is more than 150% (and can be larger). Naturally, the differences become even more pronounced when the fixed charges are not zero ($\Sigma_s \neq 0$) and when lower concentrations are considered.

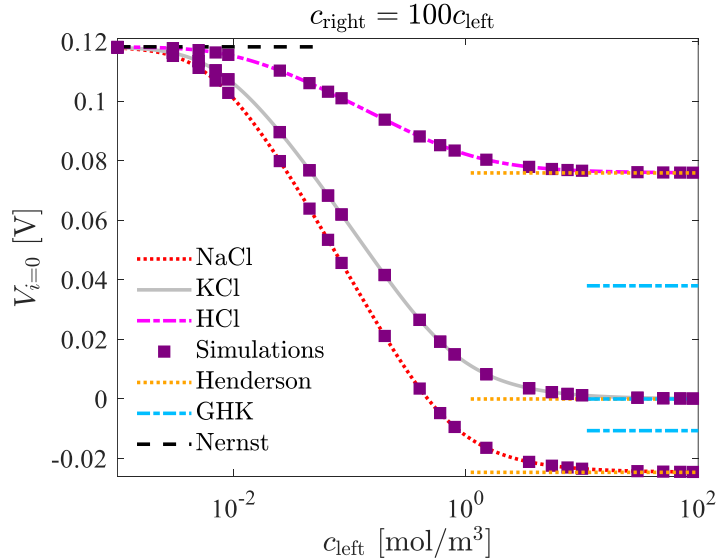

**Figure 3**. The zero current voltage, $V_{i=0}$ versus concentration for several electrolytes at a ratio $c_{\text{right}}/c_{\text{left}} = 10^2$. The RED model [Eq. (2)] shows remarkable correspondence

to numerical simulations. At high concentrations, the general solution reduces to the Henderson potential [Eq. (6)], while at low concentrations, the potential reduces to the ideal selective limit of the Nernst potential [Eq. (7)]. Except for KCl, when both the Henderson potential and the GHK potential predict zero voltage, the GHK potential [Eq. (1)] appears always to be off. Adapted and reprinted with permission from (Green, 2025a, 2025b). Copyright (2025) (American Physical Society).

## 6. Discussion

Thus far, we have demonstrated that the GHK model does not universally hold. However, such a statement is also true for the alternative ED/RED model. One needs to remember that both solutions are two different limiting solutions of the PNP equations, subject to different underlying assumptions. One should remember that, as with any assumption, there are times when the assumption holds, and there are times when the assumption fails. In this sense, both models are correct. However, since the domains of overlap of the assumptions do not generally overlap, typically, when one model is correct, the other cannot be. Hence, it is important to understand what is the criteria for each model in its limiting domain of validity.

**Surface charge effects and Debye length overlap.** There are several important differences between the models. As we have stated, one such difference is in the effects of $\Sigma_s$, wherein both models give different predictions for $\Sigma_s = 0$! The origin of this can be associated with the ratio $\Sigma_s/c_{\text{left}}$ dependence on the Debye length (for this argument, one can replace $c_{\text{left}}$ with $c_{\text{right}}$ without any loss of generality). The Debye length, $\lambda_D$, is the thin layer adjacent to a charged surface wherein the effects of the surface charge play a significant role and where, typically, the concentration of the positive and negative ions is not equal. If $\Sigma_s = 0$, in principle, there are no charges to break the symmetry of the positive and negative ions, and the concept of the Debye length becomes void, such that $\lambda_D = 0$. However, for now, as (implicitly) assumed by GHK, we will allow for $\lambda_D \neq 0$ even when $\Sigma_s = 0$.

The ratio $\Sigma_s/c_{\text{left}}$ depends on the degree of overlap of the Debye length to the channel radius, denoted by $\varepsilon_a = \lambda_D/a$, such that one can find that $\Sigma_s/c_{\text{left}} \sim \varepsilon_a^2$ (Green, 2025b). This ratio is immensely important in determining whether or not the channel is selective or not ("selectivity" is discussed further below). However, there is another equally important ratio, this is the Debye length to the channel length, $\varepsilon_L = \lambda_D/L$. This ratio represents how much the effects of the surface charge extend from one entrance to the pore to the other entrance/exit. For most pores, the ratio $L/a \gg 1$ is such that one cannot have overlap across the channel length without having radial overlap. For example, for 0.150 M (physiological scenario) and 0.11 M (life-threatening hyponatremia), the Debye length is 0.7 and 1 nm, respectively. Suppose that the channel has a radius of 2 nm but a length of 10 nm; one would have slight overlap in the radial direction but little or no overlap in the axial direction.

The difference between GHK and ED/RED is how they account for $\Sigma_s/c_{\text{left}}$ and the effects of $\varepsilon_L$. In ED/RED, the ratio $\Sigma_s/c_{\text{left}}$ is kept arbitrary (and can be considerably large), and it is assumed that $\varepsilon_L$ is considerably small. GHK uses the exact opposite assumption, wherein $\varepsilon_L$ is arbitrary (and even needs to be considerably large), while $\Sigma_s/c_{\text{left}}$ is considerably small (in fact, it is assumed to be zero). This change leads to different modelling and different results/predictions. Based on ED/RED, when $\Sigma_s = 0$ (and the self-consistent limit of $\varepsilon_a = \varepsilon_L = 0$), the positive and negative concentrations are identical, such that the electrolyte is electroneutral. Based on GHK, even when $\Sigma_s = 0$, the concentrations are not equal.

**Selectivity.** Notwithstanding the mathematical differences, there is another crucial difference between the GHK and ED/RED communities in their use of the word 'selectivity'. A detailed discussion on the two definitions is provided in Sec. II.B.1 in (Green, 2025b), and thus we discuss it here briefly. GHK defines selectivity to be a **relative ratio** of the fluxes of two different ionic species. If one species has a larger diffusion coefficient, the channel is selective towards the slower ion. For the case of two species (positive and negative), say NaCl, even though both charges are going through the system (for $\Sigma_s = 0$), Cl has a larger diffusion coefficient, and therefore the channel is selective towards potassium. It is worthwhile to note that two recent molecular dynamics simulation works (Muccio et al., 2025; Gargano et al., 2025) have yielded results that have called into question this relative definition and have suggested that it be reevaluated. The solution likely lays with ED/RED's definition for selectivity.

In ED/RED, one uses an **absolute metric**. In particular, selectivity is defined through the transport number, which is the ratio of the electric flux carried by one species relative to the total current. Similar to the previous definition, one finds that at high concentrations ($\Sigma_s = 0$) that the flux carried by Cl is larger. However, for low concentrations, when there is a high degree of Debye length overlap and $\Sigma_s/c_{\text{left}} \neq 0$, one finds the opposite result: Cl is almost entirely excluded from the channel, and now the contribution of Cl to the current is practically zero. In ED/RED, the selectivity is also associated with the channel (and its associated surface charge) and not just the ions. Thus, at high concentrations, when both species are transported, the channel is "vanishingly selective", while at low concentrations, where only one species is transported, the channel is "highly/ideally selective".

**Has the new model been compared with experiments?** A natural question arises – how well does this new theory compare to established experiments of physiological systems? Our answer to this question can be divided into several parts.

First, we have not conducted these experiments. Nor have we tried to compare the theory to the thousands of published works on this. There would be very little point in such a grand undertaking whose final outcome can already be predicted: there will be works that have better correspondence to our new theory, and there will be works that will be better described by GHK. This will not come as a surprise, since the result depends on the system parameters and experimental conditions.

Second, there is an issue of practicability, which can be divided into two. It is impossible to refute thousands of experiments, where the parameters and conditions, as well as the data analysis methods (which should include details regarding the fitting parameters but do not include these details), are not reported in sufficient enough detail. Furthermore, suppose that several, if not many, key papers for the community did include the details, a particular focus on one paper or another could be misconstrued as though it is a particular paper that is incorrect, instead of the reevaluation of the GHK methodology. Thus, we have decided not to compare our results with any experimental work and focus on the general principle.

The comparison of our new theory with new experiments should be conducted by experimentalists with new and unbiased experiments who try to fit both our model and GHK. As they do so, we recommend that the use of permeabilities be avoided altogether. Instead, one should use the models that are given here in terms of the diffusion coefficients. Such an approach ensures that there are no free fitting parameters and that the comparison of the models is on equal footing.

**Additional issues with interpreting experiments with theory.** Suppose, however, that all the required parameters and conditions were reported in experimentals. The two models presented in this work (GHK and RED) face an additional conceptual challenge, faced in every comparison of an experimental measurement to a simplified theoretical model: this is the lack of a one-to-one correspondence between the two. In the following, we will expand upon several issues that complicate the comparison.

<u>Realistic vs ideal.</u> In ideal theory, the starting point is to simplify the governing equations, boundary conditions by asserting one assumption or another (for example, "uniform electric field" or "uniform geometry") so that a final and tractable expression can be derived. In contrast, realistic biological experiments are inherently complicated, wherein it is all but impossible to recreate the ideal system. Thus, any comparison that is based on inference is always approximate and should be treated as thus.

Ideal geometry versus realistic geometry. As we have already discussed and shown in Figure 1, the idealized theories of GHK and ED/RED both assumed a quasi-1D system with a uniform and straight cross-section. In reality, an ion channel is everything but that. It is torturous with a varying cross-section and often embedded in a curved cellular membrane. This alone can introduce a non-trivial difference in the transport characteristic, including current rectification.

Single channel versus multiple channels. In patch clamp experiments, the measured current is not necessarily transported through a single channel but rather through several channels. One then has to know or assume what the pore density (or number of pores) is in order to get a 1D average of the current, which can be compared to theory. Such an approach entails an additional layer of assumptions.

Homogenous versus heterogeneous systems. In a realistic system, part of the surface is insulating, while part of it allows for the transport of current through it (Figure 1). Such a system is called a heterogenous system. In contrast, in both GHK and ED/RED models, it is assumed that the entire cross-section of the system allows for transport through it. Such a system is called homogeneous. The difference between the two systems is that a homogeneous system is typically 1D, while a heterogeneous system is typically 3D (and under certain degenerate conditions is 2D). Thus, to compare the total measure current to the theoretical current density, one needs to divide the total current by the total area or the pore area (Sebastian & Green, 2023). While either division can be appropriate, they will yield different results.

Interchannel communication. In reality, most systems are heterogeneous and include multiple channels whereby the current through one channel depends on other channels, i.e., interchannel communication. Interchannel communication requires that the more appropriate 3D approach replace the oversimplification of a 1D system. However, the transition to 3D is not without additional mathematical complications (Sebastian & Green, 2023).

Access resistance. Thus far, both GHK and ED/RED models have explicitly assumed that the effects of the reservoirs are zero. In the field of physiology, Hall (Hall, 1975) showed that what is now commonly termed access or Hall resistances can be equal to or larger than the pore resistances. In ED/RED, Hall's resistance has been extended (Yossifon & Chang, 2010; Sebastian & Green, 2023). The Hall resistance was originally derived for a single circular nanopore embedded into a membrane surrounded by infinite reservoirs with a high concentration electrolyte, such that the effects of the surface charge are negligible ($\Sigma_s/c_{\text{left}} \to 0$). In ED/RED, the equivalent resistance was derived for a nanochannel with an arbitrary shape (Sebastian & Green, 2023). Furthermore, here, the size of the reservoirs is assumed to be finite – this introduces an additional resistance which is often neglected. Furthermore, the ED/RED "access-resistance" can be extended to low concentrations, $\Sigma_s/c_{\text{left}} \neq 0$, where GHK does not hold.

Estimating the fixed charges ($\Sigma_s$). GHK explicitly assumes that $\Sigma_s = 0$. In reality, this is never the case. In fact, one can say with almost absolute certainty that $\Sigma_s/c_{\text{left}} \neq 0$. However, estimating $\Sigma_s$ can now be considered a routine and rather trivial experimental procedure. One needs to measure the electrical conductance of a system at symmetric concentrations ($c_{\text{left}} = c_{\text{right}}$). Then, $\Sigma_s$ can be curve-fitted. See (Green, 2025b) for more details.

At the end of the comparison and interpretation process, one needs to remember that the goal of the simplified theoretical framework is to provide a way to understand the highly complicated experiments. The ultimate goal, in our view, is to capture the **essence** of the physics and to determine which variables contribute significantly and which may be taken as negligible.

## 7. Conclusions

In this short report, we have presented two models that are derived from the same mathematical formulation (Appendix A). Each model is derived using a slightly different approach (Appendix B). In particular, the two models use two different assumptions: "uniform field" versus "electroneutrality". This "small" difference leads to rather drastic changes.

Importantly, we have demonstrated that the GHK model is not as universal as previously thought, and it is not the only unique solution to the governing place. We present an equally robust model that leverages all the theoretical modeling capabilities developed independently by the ED/RED communities. This new model shows remarkable correspondence to non-approximated numerical simulations.

In this work, we have discussed the new ED/RED model only in terms of the resting potential of the membrane ($V_{i=0}$). However, the ED/RED model has additional predictions for the Ohmic conductance, the transport number, the current at zero voltage ($i_{V=0}$), and more. These issues are discussed in our two recently published works (Green, 2025a, 2025b).

We believe that this work, which highlights the differences between the electrophysiology community and the desalination and energy harvesting communities, can also serve as a bridge to unify the methods, approaches, and understanding of these two communities.

## Additional information

**Acknowledgments.** We thank Prof. James Butler for their careful reading of this letter and the ensuing discussions.

**Data availability statement.** Data sharing is not applicable to this article as no new data were created or analyzed in this study.

**Competing interests.** The authors declare that they have no conflicts of interest.

**Funding.** This work was supported by Israel Science Foundation grants 204/25. We acknowledge the support of the Ilse Katz Institute for Nanoscale Science and Technology and the Pearlstone Center for Aeronautical Engineering Studies.

## Appendix A: Governing equations

The steady-state convection-less 1D Poisson-Nernst-Planck equations for a two-species electrolyte through a charged nanochannel are

$$\varepsilon_0 \varepsilon_r \partial_{xx} \phi = -F(z_+ c_+ - z_- c_- - \Sigma_s). \tag{8}$$

$$\partial_x j_\pm = -\partial_x \left( D_\pm \partial_x c_\pm \pm \frac{z_\pm D_\pm F}{RT} c_\pm \partial_x \phi \right) = 0, \tag{9}$$

Here, $z_\pm = \pm z$ are the ionic valences (which are assumed to be equal in the absolute sense), and $\varepsilon_0$ and $\varepsilon_r$ are the permittivity of free space and the relative permittivity, respectively. The Poisson equation [Eq. (8)] governs the electrical potential, $\phi$, while the Nernst-Planck equations [Eq. (9)] govern ion flux conservation for a positive and negative species, $c_+$ and $c_-$, respectively, with the appropriate fluxes $j_\pm$. The space charge density [i.e., the right side of Eq. (8)] depends on the positive and negative ions, and the fixed charges $\Sigma_s$. At the two ends of the system, we have two different bulk concentrations and a total potential drop, $V$

$$\begin{aligned} \phi(x=0) = V, \quad c_\pm(x=0) = c_{\text{left}} \\ \phi(x=L) = 0, \quad c_\pm(x=L) = c_{\text{right}} \end{aligned}. \tag{10}$$

Equations (8)-(10) are the **exact** same equations solved in both seminal GHK papers for two species.

In principle, this formulation can be extended to an arbitrary number of charged species. Then, Eq. (10) for $c_{\text{left}}$ and $c_{\text{right}}$ needs to be modified appropriately to account for the additional species and bulk electroneutrality. However, for two charged species (one positive and one negative), wherein one must have that the total net concentration is electroneutral, one must require, as in Eq. (10), that at $x$=0, $L$, the positive and negative concentrations are the same (such that $c_+ = c_-$).

Here, we focus on just two species for three reasons. First, it is much simpler, and importantly, less abstract. Second, thus far in the alternative formulation, we have been able to derive a solution only for two species, which stands in contrast to GHK, which has been derived for an arbitrary number of species. Third, the breakdown in the universality of GHK for two species necessarily implies that the breakdown occurs for an arbitrary number of species. Hence, a model with two species is sufficient.

In this work, we present numerical simulations of Eqs. (8)-(10) for the case of a sufficiently long channel, wherein $\varepsilon_L = \lambda_D/L \ll 1$. In numerical simulations, $\Sigma_s$ and the concentration ratio and $c_{\text{right}}/c_{\text{left}}$ are set, while $c_{\text{left}}$ is varied over six decades of concentrations. Details of the numerical simulations can be found in the supplementary material of (Green, 2025b).

**Appendix B: Solution of the Poisson equation**

The PNP equations are a non-linear set of coupled differential equations. In 2D and 3D, they are partial differential equations. In 1D, they are simplified to ordinary differential equations. The solution of these equations is non-trivial and typically requires an assumption of one form or another for simplification. Here, this assumption is applied to the Poisson equation [Eq. (8)].

The approach of Goldman (Goldman, 1943) and Hodgkin-Katz (Hodgkin & Katz, 1949) is to assert that the effects of the space charge density [the right-hand side of Eq. (8)] are negligible, such that the Poisson equation can be replaced by the Laplace equation, $\partial_{xx}\phi=0$. The solution to this equation is a linear profile with a uniform/constant electric field. However, thereafter, electroneutrality (local and/or global) is not satisfied.

The electroneutral model of ED/RED takes the exact opposite approach. It is assumed that the effects of the Laplacian operator are negligible, such that Eq. (8) satisfies electroneutrality, and the right-hand side of Eq. (8) can be written as $c_+ = c_- + \Sigma_s/z$. However, thereafter, the non-linear Nernst-Planck equations need to be solved for the remaining unknown fields (including the electrical potential, which cannot be calculated explicitly).

Unsurprisingly, both approaches, which utilize different assumptions, yield different results. Each is appropriate for different scenarios (i.e., different parameters).